Review                                                                                                           RRJoST

# Two Decades of Research at the University of Lagos (2004-2023): A Scientometric Analysis of Productivity, Collaboration, and Impact

Muneer Ahmad[1,]*, Samuel Ibor Ubi[2]

*Abstract*
*This paper presents a scientometric analysis of research output from the University of Lagos, focusing on the two decades spanning 2004 to 2023. Using bibliometric data retrieved from the Web of Science, we examine trends in publication volume, collaboration patterns, citation impact, and the most prolific authors, departments, and research domains at the university. The study reveals a consistent increase in research productivity, with the highest publication output recorded in 2023. Health Sciences, Engineering, and Social Sciences are identified as dominant fields, reflecting the university's interdisciplinary research strengths. Collaborative efforts, both locally and internationally, show a positive correlation with higher citation impact, with the United States and the United Kingdom being the leading international collaborators. Notably, open-access publications account for a significant portion of the university's research output, enhancing visibility and citation rates. The findings offer valuable insights into the university's research performance over the past two decades, providing a foundation for strategic planning and policy formulation to foster research excellence and global impact.*

**Keywords:** Research productivity, University of Lagos, Citation Impact, Collaboration patterns, Research trends, Higher education research, Interdisciplinary research.

## INTRODUCTION

The University of Lagos, commonly referred to as UNILAG, is one of the foremost institutions of higher learning in Nigeria. Established in 1962, it has grown to become a major center of academic excellence and a leading contributor to research and innovation in the country (University of Lagos, 2023).[1] The university is located in Akoka, Yaba, Lagos, which is one of the most vibrant and diverse cities in Nigeria. This strategic location has significantly influenced the university's development, providing students and staff with a rich cultural and professional environment.

*Author for Correspondence
Muneer Ahmad
E-mail: muneerbangroo@gmail.com

[1]Chief University Librarian, The Iddi Basajjabalaba Memorial Library, Kampala International University, Kansanga, Uganda.
[2]Deputy Librarian, The Iddi Basajjabalaba Memorial Library, Kampala International University, Kansanga, Uganda

UNILAG offers a wide range of undergraduate and postgraduate programs across various fields, including sciences, engineering, arts, social sciences, law, and business. With over 50,000 students, it is one of the largest universities in Nigeria. The institution prides itself on its diverse student body and faculty, which includes individuals from different parts of Nigeria and beyond (National Universities Commission, 2023). [2] This diversity fosters a rich learning environment and prepares students for global competitiveness.

                21

The university's infrastructure and facilities are also noteworthy. UNILAG boasts state-of-the-art laboratories, libraries, and lecture halls that support its academic programs. The institution's library, for instance, is one of the best in Nigeria, offering a vast collection of books, journals, and electronic resources to support research and learning [3] (University of Lagos Library, 2023). Additionally he university has modern sports facilities, residential halls, and health services that cater to the needs of its students and staff.

Research and innovation are at the core of UNILAG's mission. The university has numerous research centers and institutes that focus on addressing key national and global challenges. For example, the University of Lagos Research and Innovation Office (R&I) coordinates research activities and promotes collaboration with industries and other academic institutions[4] (University of Lagos R&I, 2023). This emphasis on research has led to significant breakthroughs in various fields, including science and technology, healthcare, and social sciences.

The university also places a strong emphasis on entrepreneurship and skills development. Through its Centre for Entrepreneurship and Skills Development (CESD), UNILAG provides students with the skills and knowledge needed to succeed in the competitive job market. The center offers various programs and workshops that encourage innovation and self-reliance among students[5] (University of Lagos CESD, 2023).This initiative aligns with the university's goal of producing graduates who are not only academically sound but also capable of contributing to the economic development of Nigeria.

UNILAG's alumni network is another testament to its impact on society. Graduates of the University of Lagos have gone on to excel in various fields, including politics, business, academia, and entertainment. Notable alumni include Nigeria's former president, Olusegun Obasanjo, and the renowned writer, Chimamanda Ngozi Adichie[6] (University of Lagos Alumni Association, 2023). The university's strong alumni network provides current students with valuable mentorship and networking opportunities.

In conclusion, the University of Lagos stands out as a beacon of academic excellence and innovation in Nigeria. Its diverse programs, state-of-the-art facilities, and strong emphasis on research and entrepreneurship make it a leading institution in the country. With its strategic location in Lagos, the university continues to attract students and scholars from around the world, contributing significantly to the socio-economic development of Nigeria.

**REVIEW OF LITERATURE**
Scientometric studies have become an essential tool in evaluating research performance and academic output. These studies utilize various quantitative metrics to analyze publications, citations, and other research indicators. The importance of scientometric indicators in the academic landscape has been extensively documented, providing insights into research productivity, collaboration patterns, and the impact of open-access publishing. discusses the significance of scientometric indicators such as citation counts, h-index, and journal impact factors in assessing the research performance of universities. These metrics provide quantitative measures that facilitate the comparison of research outputs across institutions. Moed[7] emphasizes the need for a balanced evaluation, combining both quantitative and qualitative assessments.

titled "Lotka's Law and Authorship Distribution in Coronary Artery Disease Research in South Africa", [8]examines the applicability of Lotka's Law to the authorship patterns in this field. Using publication data from the Web of Science for the period 1990-2019, the study analyzes 1,284 research papers authored by South African scientists. It applies bibliometric tools and statistical tests, including the Kolmogorov-Smirnov test, to assess the fit of Lotka's Law. The findings reveal that while the general distribution of authorship follows Lotka's inverse square law, the specific parameters for South African CAD research do not perfectly align, highlighting deviations in author productivity. The study underscores the importance of localized analysis in bibliometrics.

The work of Bornmann, de Moya Anegón, and Leydesdorff (2012) [9] explores the application of bibliometric methods in university rankings. They highlight how citation-based metrics and collaborative research networks influence the positions of universities in global rankings. The authors argue for the inclusion of diverse bibliometric indicators to ensure a comprehensive evaluation.

Larivière, Ni, Gingras, Cronin, and Sugimoto (2013) [10] analyze gender disparities in scientific publishing within universities. Their scientometric study reveals significant differences in publication rates, citation counts, and collaborative patterns between male and female researchers. The study calls for institutional policies to address these disparities.

Katz and Martin (1997) [11] investigate the impact of inter-university collaborations on research output. Their scientometric analysis shows that collaborative research tends to receive higher citation rates compared to non-collaborative work. The study suggests fostering collaborative networks to enhance research impact.

Abramo, D'Angelo, and Cicero (2012) [12] discuss methodologies for assessing the research productivity of universities using scientometric indicators. They focus on field-normalized citation metrics to account for disciplinary differences in publication and citation behaviors. Their study underscores the importance of context-sensitive evaluations.

Pouris and Pouris (2009) [13] conduct a scientometric analysis of universities in emerging economies, particularly in Africa. They highlight the challenges these institutions face in achieving global research visibility and impact. The study suggests strategic investments in research infrastructure and international collaborations.

Gargouri, Larivière, Gingras, Carr, and Harnad (2010) [14] examine the impact of open-access publishing on university research's visibility and citation rates. Their scientometric study indicates that open-access articles receive more citations than those behind paywalls. The authors advocate for institutional policies supporting open access.

Bonaccorsi, Daraio, Lepori, and Slipersaeter [15] analyze the distribution of scientific excellence across European universities using scientometric data. They identify key factors that contribute to high research performance, including funding, infrastructure, and international collaborations. The study provides policy recommendations to enhance scientific output.

Geuna and Martin (2003) [16] explore the relationship between research funding and university outputs using scientometric methods. Their study finds a positive correlation between funding levels and research productivity, measured in terms of publications and citations. The authors discuss the implications for research policy and funding allocation.

Porter and Rafols (2009) [17] investigate the prevalence and impact of interdisciplinary research within universities. Their scientometric analysis shows that interdisciplinary studies often receive higher citation rates and have a broader societal impact. The authors recommend fostering interdisciplinary collaborations to enhance research innovation.

Ahmad (2022) [18] conducted a comprehensive scientometric investigation on coronary artery disease research within BRICS countries, covering the period from 1990 to 2019. The study revealed a consistent and rapid growth in research output over the three decades, reflecting the increasing global focus on coronary artery disease. The study also highlighted a significant rise in both country-wise and year-wise collaborative research efforts, suggesting that international collaboration plays a key role in driving research productivity in this field. The author found that journal articles were the preferred medium for disseminating research findings, underlining the prominence of scholarly journals in cardiovascular research. Among the BRICS nations, the People's Republic of China was the top contributor to coronary artery disease research, while South Africa had the lowest publication output. This disparity underscores the varying levels of research investment and focus within the BRICS bloc. English emerged as the dominant language for scientific communication, reinforcing its role as the lingua franca of global research. The study also identified a trend toward multiple authorship patterns, which were more prevalent than single-author publications. This finding aligns with the broader trend in scientific research where collaboration, often on a large scale, enhances both the quality and quantity of research output. The research work contributes valuable insights into the research dynamics of coronary artery disease within BRICS countries, offering a basis for understanding how global collaboration and knowledge dissemination are shaping this critical area of medical research.

**OBJECTIVES**
- To look at the research output growth pattern of the University of Lagos year over year.
- Discover which of the University of Lagos authors are the most prolific.
- Determine which esteemed journals publish academic works from the University of Lagos.
- Determine which nations are best suited to work together to publish the findings of their research. The aim is to map the top 20 authors, collaborating countries, and institutions according to the number of research papers they have written to determine publication density.

**METHODOLOGY**
The current study utilizes the Clarivate Analytics Web of Science (WOS) database to quantify the research productivity of the University of Lagos. Using a basic search, information was obtained from the WOS database By searching for the term "University of Lagos" in the affiliation field, with the date span set to 2004–2023 and the indexes SCI-EXPANDED, SSCI, and AHCI, publications associated with the University of Lagos were identified. A total of 4852 articles were located.

To ensure data quality and consistency, several measures were taken. First, the search parameters were carefully defined to minimize the risk of retrieving irrelevant or duplicate records. Only publications explicitly affiliated with the University of Lagos were included, and the date span was chosen to provide a comprehensive overview of the institution's research output over nearly two decades. The data extracted from WOS was further cleaned and verified for consistency by cross-referencing with institutional records where available.

The analysis was conducted using bibliometric tools and software, including Histcite, RStudio, and MS Excel, which allowed for detailed examination and visualization of the data. These tools were selected for their ability to handle large datasets and perform complex bibliometric analyses efficiently.

Despite these measures, the study acknowledges certain limitations inherent in using the Web of Science database, particularly in the context of African institutions. One significant limitation is the potential underrepresentation of research outputs that are published in local or regional journals, which may not be indexed by WOS. This can result in an incomplete picture of an institution's total research productivity. Additionally, there may be discrepancies in the affiliation information provided in WOS, which could lead to either underreporting or overreporting of publications attributed to the University of Lagos.

Recognizing these limitations, the findings of this study should be interpreted with caution, and further research incorporating additional databases and local repositories is recommended to provide a more holistic view of research productivity.

## RESULTS AND DISCUSSION
### General Data Overview

The study covers 20 years of data from various publications. A total of 1601 sources were analyzed, including a combination of journals, books, and other types of publications. A substantial 4852 documents were included in the analysis. The slight negative growth rate suggests that there was a minor decrease in the number of documents published annually over the 20-year period, indicating either saturation in the field or a decline in publication output in recent years. On average, the documents in the study are approximately 7.7 years old, indicating that the analyzed works are relatively recent. Each document, on average, has been cited around 28 times, indicating a high level of influence or impact of the documents in this dataset. A massive number of references (146,190) were used, reflecting the extensive citation network and the scholarly depth of the documents. This represents the number of keywords or key concepts assigned to documents through algorithms that help capture essential topics. A total of 10,929 unique keywords provided by the authors were analyzed, reflecting the diversity of topics and areas covered by the documents. A total of 28,472 authors contributed to the documents in this dataset, indicating a large research community. Only 164 authors contributed to single-authored papers, reflecting that most works are collaborative. Out of 4852 documents, only 239 were single-authored, further reinforcing the collaborative nature of the field. On average, each document had 16.9 co-authors, which suggests that the field involves a high degree of teamwork and collaboration across research projects. Nearly 48% of the documents were co-authored internationally, indicating strong global collaboration in the research field. The table 1. provides a comprehensive summary of research activity and collaboration in this study, offering valuable insights into the trends and influence of the body of work analyzed.

### Evaluation of the Annual Output of Publications of the University of Lagos

The table 2 provides a detailed breakdown of publications and citations over the years from 2004 to 2023, showing the growth and citation impact of research output annually. The number of publications steadily increased over the years, peaking in 2023 with 498 documents (10.26% of the total). Between 2004 and 2023, there's clear growth in research output, with earlier years (2004-2010) having fewer publications, while recent years (2016–2023) saw significantly higher numbers. The years 2018-2023 contribute the most significant percentage of records (with each year exceeding 6% of the total records), indicating a substantial increase in research output in the last five years. The highest global citation impact comes from the 2018 publications (25,463 citations, 18.73% of total global citations). This suggests that research published in 2018 had a massive influence on the global scholarly community. Local citations (citations within the dataset) also peaked in 2018, with 282 citations (9.99% of total local citations). Notably, the year 2023 has the highest number of records (498 records, 10.26%) but received the lowest local citation impact, with only 36 citations (1.28%). This is understandable, as research from 2023 may not have had enough time to accumulate citations.

**Table 1.** Main information about data.

| S.N. | Description | Results |
|---|---|---|
| 1 | Timespan | 2004:2024 |
| 2 | Sources (Journals, Books, etc.) | 1601 |
| 3 | Documents | 4852 |
| 4 | Annual growth rate % | -1.73 |
| 5 | Document average age | 7.71 |
| 6 | Average citations per doc | 28.01 |
| 7 | References | 146190 |

|    | *Document contents*            |       |
|----|-------------------------------|-------|
| 8  | Keywords Plus (ID)            | 8769  |
| 9  | Author's Keywords (DE)        | 10929 |
|    | *Authors*                     |       |
| 10 | Authors                       | 28472 |
| 11 | Authors of single-authored docs | 164 |
|    | *Authors collaboration*       |       |
| 12 | Single-authored docs          | 239   |
| 13 | Co-authors per doc            | 16.9  |
| 14 | International co-authorships % | 48.02 |

**Table 2.** Annual distribution of publications and citations.

| S.N. | Year | Records | %     | Rank | TLCS | %     | TGCS   | %      |
|------|------|---------|-------|------|------|-------|--------|--------|
| 1    | 2004 | 61      | 1.26  | 20   | 72   | 2.55  | 1190   | 0.88   |
| 2    | 2005 | 88      | 1.81  | 19   | 73   | 2.59  | 1693   | 1.25   |
| 3    | 2006 | 95      | 1.96  | 18   | 86   | 3.05  | 2988   | 2.20   |
| 4    | 2007 | 142     | 2.93  | 17   | 156  | 5.53  | 2853   | 2.10   |
| 5    | 2008 | 185     | 3.81  | 13   | 123  | 4.36  | 3603   | 2.65   |
| 6    | 2009 | 171     | 3.52  | 16   | 113  | 4.00  | 3304   | 2.43   |
| 7    | 2010 | 184     | 3.79  | 14   | 113  | 4.00  | 2712   | 2.00   |
| 8    | 2011 | 186     | 3.83  | 12   | 108  | 3.83  | 2658   | 1.96   |
| 9    | 2012 | 205     | 4.23  | 10   | 107  | 3.79  | 3200   | 2.35   |
| 10   | 2013 | 176     | 3.63  | 15   | 108  | 3.83  | 2778   | 2.04   |
| 11   | 2014 | 194     | 4.00  | 11   | 123  | 4.36  | 2504   | 1.84   |
| 12   | 2015 | 217     | 4.47  | 9    | 81   | 2.87  | 2973   | 2.19   |
| 13   | 2016 | 252     | 5.19  | 8    | 148  | 5.24  | 4208   | 3.10   |
| 14   | 2017 | 262     | 5.40  | 7    | 248  | 8.78  | 13413  | 9.87   |
| 15   | 2018 | 320     | 6.60  | 5    | 282  | 9.99  | 25463  | 18.73  |
| 16   | 2019 | 314     | 6.47  | 6    | 205  | 7.26  | 19406  | 14.28  |
| 17   | 2020 | 377     | 7.77  | 4    | 205  | 7.26  | 19618  | 14.43  |
| 18   | 2021 | 444     | 9.15  | 3    | 220  | 7.79  | 12295  | 9.05   |
| 19   | 2022 | 481     | 9.91  | 2    | 216  | 7.65  | 6643   | 4.89   |
| 20   | 2023 | 498     | 10.26 | 1    | 36   | 1.28  | 2415   | 1.78   |
| *Total* |   | *4852*  | *100.00* |   | *2823* | *100.00* | *135917* | *100.00* |

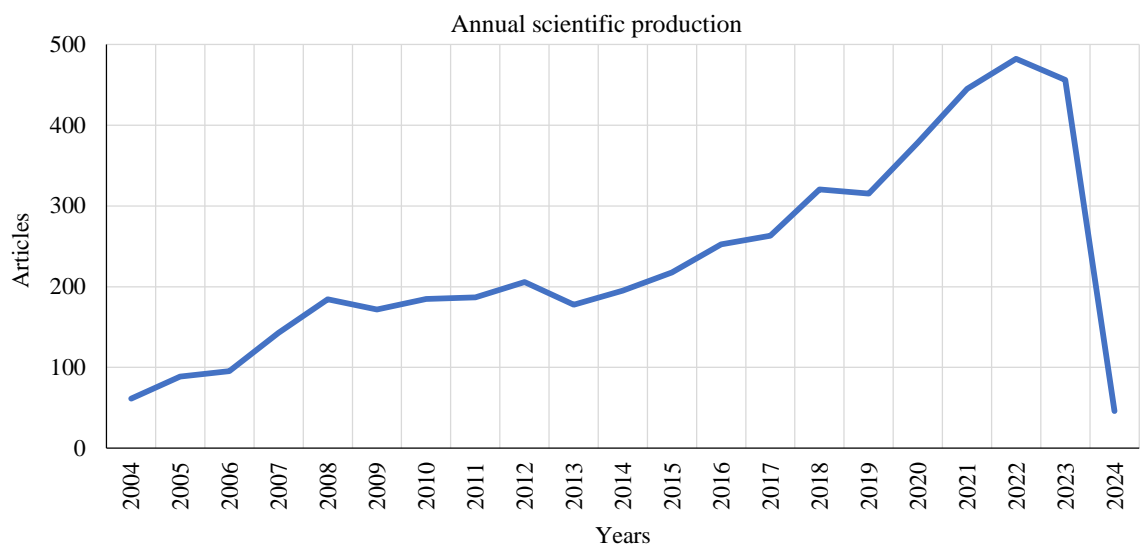

**Figure 1.** Annual scientific production (Source: web of science)

In contrast, older years, such as 2017 and 2018, have fewer records but a much higher percentage of citations, showing their continued influence in the academic community. In Figure 1 2023 ranks 1st in terms of the number of records (498 documents), followed by 2022 with 481 documents. However, the most impactful years in terms of global citations are 2018, 2019, and 2020, with each having a significant share of global citations (2018 with 18.73%, 2019 with 14.28%, and 2020 with 14.43%). The earlier years (2004–2010) have a lower contribution in terms of records, with percentages ranging from 1.26% to 3.79%. However, the impact of citations is still significant. For example, the publications from 2007 have the highest global citation score among the earlier years, with 2853 citations and a 2.10% contribution to global citations. The global citation score peaked in 2018 with 25,463 citations (18.73%), showing that publications from that year were highly influential. However, as we approach the more recent years (especially 2023), the total global citation score decreases, reflecting the fact that newer publications generally take time to accumulate citations. The table 3 indicates a clear growth in research output, particularly in the last decade. The number of records has increased steadily, with the highest number of publications in 2023 and 2022. 2018 stands out as a particularly impactful year both in terms of local and global citations despite having fewer records compared to 2023. In recent years, like 2023, while contributing the most records, there have been fewer citations, as citations typically accumulate over time. This data offers a snapshot of the evolution and influence of research output over the two decades, showing a trend of increasing publications and fluctuating citation impact across different years.

**Analysis of the Publication Output of Top 20 Authors of the University of Lagos**

The table 3 ranks the top 20 authors from the University of Lagos, Nigeria, based on their h-index, publication output, and citation scores. The h-index reflects both the productivity and the impact of the authors' publications, while citation metrics provide insight into the influence of their work. Mokdad AH has the highest h-index of 71, making him the most prolific and impactful author in terms of publication and citation count. He has received 63,780 citations within his h-core and 64,748 citations across all 97 articles. Olagunju AT follows closely with an h-index of 69 and a significant 61,999 citations within his h-core. His total citation count is 63,302, based on 132 published articles, making him both highly productive and influential. Gupta R and Haj-Mirzaian A stand out with impressive citation totals of 81,138 and 81,049, respectively, despite their relatively lower h-indices of 60 and 61. These figures suggest that their publications are widely recognized and referenced in the scholarly community. Authors such as Murray CJL (h-index of 63, 60,278 citations) and Tran BX (h-index of 63, 61,089 citations) also demonstrate considerable citation influence in their fields. Authors like Samy

AM (h-index of 63) and Shaikh MA (h-index of 64) balance a high number of publications (87 and 85 articles, respectively) with strong citation numbers within their h-core (55,840 and 58,860 citations). Olagunju AT (132 articles) is one of the most prolific authors on this list, indicating not only impact but also high productivity.

**Table 3.** Publication output of top 20 authors and citation score.

| S.N. | h-index | Authors | Citation sum within h-core | All citations | All articles |
|---|---|---|---|---|---|
| 1 | 71 | Mokdad AH | 63780 | 64748 | 97 |
| 2 | 69 | Olagunju AT | 61999 | 63302 | 132 |
| 3 | 67 | Fischer F | 61916 | 62676 | 88 |
| 4 | 64 | Shaikh MA | 58860 | 59590 | 85 |
| 5 | 64 | Yonemoto N | 61884 | 62595 | 84 |
| 6 | 63 | Murray CJL | 59710 | 60278 | 78 |
| 7 | 63 | Tran BX | 60634 | 61089 | 75 |
| 8 | 63 | Samy AM | 55840 | 56630 | 87 |
| 9 | 61 | Haj-Mirzaian A | 81049 | 81725 | 79 |
| 10 | 60 | Gupta R | 81138 | 81470 | 70 |
| 11 | 60 | Mohammed S | 53333 | 54095 | 87 |
| 12 | 59 | Hay SI | 58031 | 58534 | 75 |
| 13 | 59 | Koyanagi A | 54607 | 55047 | 73 |
| 14 | 58 | Dandona L | 53151 | 53433 | 69 |
| 15 | 58 | Dandona R | 53151 | 53484 | 70 |
| 16 | 57 | Singh JA | 50889 | 51410 | 75 |
| 17 | 57 | Khan EA | 56081 | 56519 | 72 |
| 18 | 57 | Rawaf S | 53805 | 54365 | 77 |
| 19 | 57 | Kisa A | 56194 | 56766 | 74 |
| 20 | 56 | Khader YS | 52803 | 53251 | 71 |

The citation impact varies significantly across authors. For instance, Mohammed S has accumulated 53,333 citations within the h-core and 54,095 citations across 87 articles, showing consistent influence. Singh JA, despite having a slightly lower h-index of 57, has an impressive 50,889 citations within the h-core and 51,410 citations in total for his 75 articles, reflecting solid influence in his research area. The authors on this list have a strong balance between the number of articles they produce and the number of citations those articles receive, with h-indices ranging from 56 to 71. This suggests that the University of Lagos has a robust research presence with authors who are both highly productive and influential in their respective fields. While Mokdad AH leads in terms of h-index and total citations, other authors such as Olagunju AT, Fischer F, and Gupta R also make significant contributions in terms of both productivity and citation impact. The top 20 authors from the University of Lagos, Nigeria, have made a significant impact on their respective fields, as evidenced by their high h-indices and citation totals. These authors not only demonstrate high productivity (in terms of the number of articles published) but also substantial influence, as seen from their high citation counts both within and beyond their h-core. Their contributions indicate that the university is fostering high-caliber research with global recognition.

**Analysis of Source-Wise Distribution of Documents**

This table 4 provides an overview of the top 20 academic journals and sources frequently utilized by University of Lagos professors for publishing their research. It displays metrics such as the number of documents published, h-index, g-index, m-index, total citations (TC), and the year the sources started being used (PY Start). Lancet is the most utilized journal, with 36 documents and a high h-index of 45, making it the most influential source in terms of citation impact (40,650 total citations). It has a g-index of 1.714, reflecting the high citation concentration of these articles. Environmental Science and Pollution Research comes second, with 25 documents and an h-index of 48. It has received a total of 2,358 citations and has been in use since 2014, showing a strong presence in the environmental sciences for the University of Lagos professors. Journal of Ethnopharmacology, with 22 documents, has an h-index of 34 and a total citation count of 1,325. It is one of the long-standing journals utilized since 2004. African Journals such as the African Journal of Biotechnology, African Health Sciences, and Nigerian Journal of Clinical Practice are important for research in the region. These journals show lower citation counts than global journals but are significant within the African academic landscape. PLOS One is a widely used open-access journal with 18 documents, an h-index of 29, and 1,036 total citations, making it another popular publication venue for research dissemination. Figure 2 Lancet Journals are well-represented, with Lancet Infectious Diseases, Lancet Neurology, Lancet Gastroenterology and Hepatology, and Lancet Global Health all appearing in the list. These journals tend to have strong citation impacts due to their global reach and specialization in public health, infectious diseases, and clinical research. Heliyon, a relatively new journal (with publications starting in 2019), has an h-index of 17 and a high m-index of 1.833, indicating consistent and influential research output from the University of Lagos professors in this journal. Malaria Journal and Advances in Space Research also make the list with moderate citation counts, indicating research interests in both infectious diseases and space research.

**Table 4.** Top 20 source wise distribution of documents.

| S.N. | Sources | Documents | h_index | g_index | m_index | TC | PY Start |
|---|---|---|---|---|---|---|---|
| 1 | Lancet | 36 | 45 | 1.714 | 40650 | 45 | 2004 |
| 2 | Environmental Science and Pollution Research | 25 | 48 | 2.273 | 2358 | 57 | 2014 |
| 3 | Journal of Ethnopharmacology | 22 | 34 | 1.048 | 1325 | 56 | 2004 |
| 4 | African Journal of Biotechnology | 18 | 26 | 0.9 | 990 | 76 | 2005 |
| 5 | PLOS One | 18 | 29 | 1.059 | 1036 | 72 | 2008 |
| 6 | Environmental Monitoring and Assessment | 14 | 21 | 0.778 | 488 | 35 | 2007 |
| 7 | Nigerian Journal of Clinical Practice | 14 | 20 | 0.824 | 706 | 102 | 2008 |
| 8 | Lancet Infectious Diseases | 13 | 14 | 0.929 | 3398 | 14 | 2011 |
| 9 | Malaria Journal | 13 | 20 | 0.722 | 449 | 20 | 2007 |
| 10 | Advances in Space Research | 12 | 19 | 0.706 | 457 | 38 | 2008 |
| 11 | African Health Sciences | 12 | 19 | 0.706 | 510 | 63 | 2008 |
| 12 | Cochrane Database of Systematic Reviews | 12 | 22 | 0.632 | 825 | 22 | 2006 |
| 13 | Lancet Neurology | 12 | 15 | 1.714 | 14697 | 15 | 2018 |
| 14 | Chemosphere | 11 | 14 | 0.55 | 499 | 14 | 2005 |
| 15 | Heliyon | 11 | 17 | 1.833 | 394 | 54 | 2019 |
| 16 | Journal of Geophysical Research-Space Physics | 11 | 16 | 0.611 | 287 | 20 | 2007 |
| 17 | Lancet Gastroenterology and Hepatology | 11 | 12 | 1.375 | 3451 | 12 | 2017 |
| 18 | Lancet Global Health | 11 | 15 | 1.1 | 1969 | 15 | 2015 |
| 19 | Fitoterapia | 10 | 13 | 0.476 | 381 | 13 | 2004 |
| 20 | Journal of Atmospheric and Solar-Terrestrial Physics | 10 | 13 | 0.625 | 187 | 15 | 2009 |

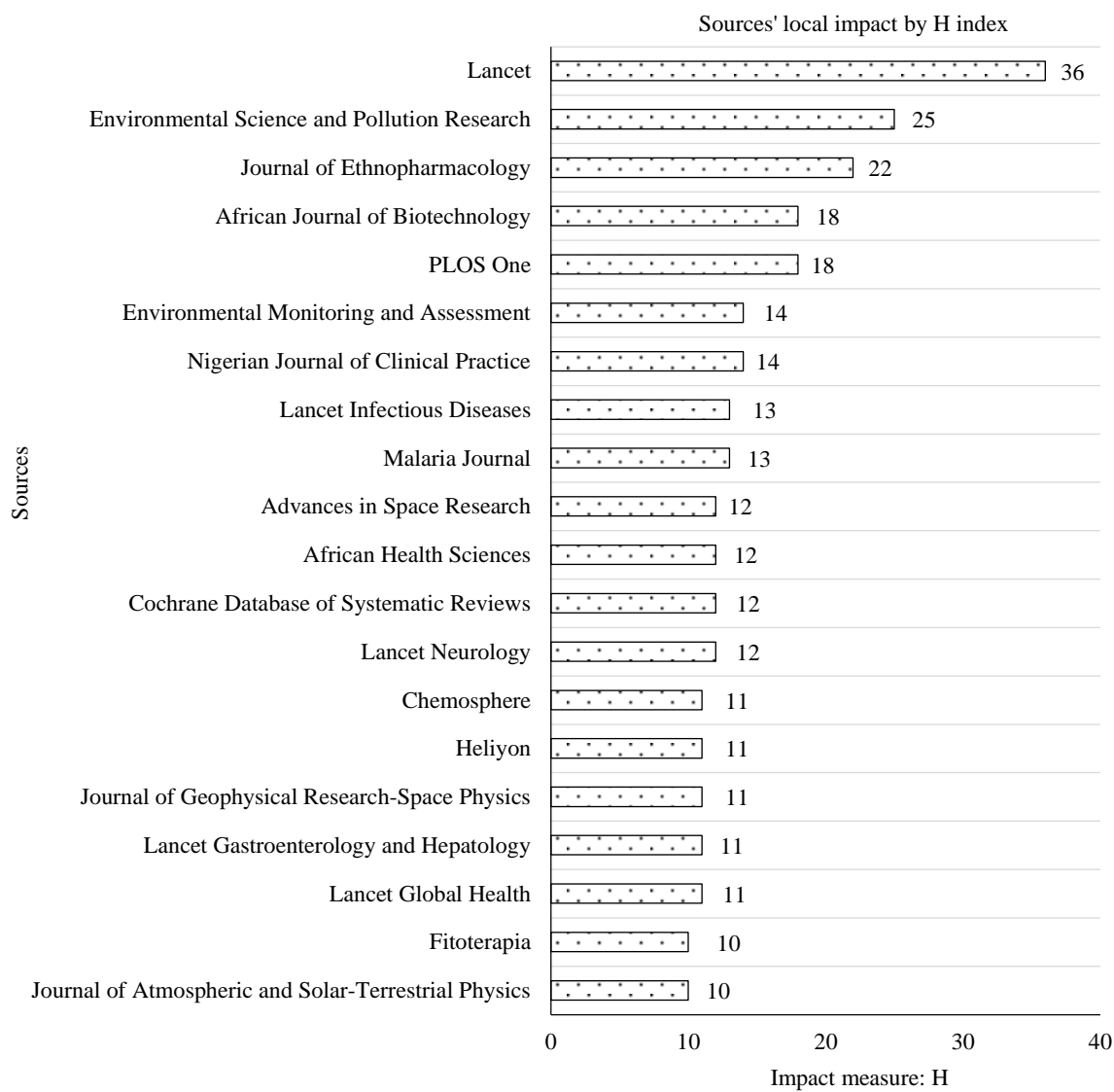

**Figure 2.** Source's Local Impact by H index (Source: web of science)

The presence of globally renowned journals like Lancet, PLOS One, and Environmental Science and Pollution Research shows that University of Lagos professors aim for high-impact publications to maximize research visibility and influence. African journals remain important for regionally focused research and dissemination, indicating a balance between global and local research priorities. Newer journals such as Heliyon and specialized journals like the Journal of Geophysical Research-Space Physics showcase the diversity in research areas covered by the professors, ranging from environmental sciences to clinical research, space physics, and public health. In summary, the top 20 preferred sources reflect a combination of high-impact global journals and key African-based outlets. Professors from the University of Lagos prioritize publishing in reputable sources that offer visibility, citations, and scholarly impact across various research fields.

**Analysis of Collaborative Ranking of Institutions**

This table 5 lists the top 20 institutions that have collaborated with the University of Lagos in research publications. It provides details such as the number of collaborative records, the total local citation score (TLCS), the total global citation score (TGCS), and the average citations per paper (ACPP) for each institution. The University of Ibadan leads in collaborations with the University of Lagos, having contributed to 359 publications, which is 7.4% of the total. It has a strong global presence with 61,115 global citations and an ACPP of 170.24, showing significant impact. Lagos State University follows closely with 355 collaborative records, representing 7.3% of the total. However, its global citation count (6,633) and ACPP (18.68) indicate a more localized impact compared to other institutions. Obafemi Awolowo University has 208 collaborative records, with a notable TGCS of 23,575 and an ACPP of 113.34, suggesting its contributions are well-cited internationally. The University of Ilorin and the University of Nigeria both have 165 collaborative publications, but the latter stands out with a higher TGCS of 11,253 and an ACPP of 68.20, compared to the University of Ilorin's 3,382 TGCS and 20.50 ACPP. Lagos University Teaching Hospital and the Nigerian Institute of Medical Research are prominent local collaborators, contributing to medical and healthcare-related research with modest citation scores and ACPP values. Ahmadu Bello University has 153 collaborative publications and a very high ACPP of 380.16, suggesting its research with the University of Lagos is highly impactful globally (58,165 TGCS). The University of Cape Town, UCL (University College London), and the University of Washington are key international collaborators, contributing to highly cited research.

**Table 5.** Ranking of collaborative institutions.

| S.N. | Institution | Records | % | TLCS | TGCS | ACPP |
|---|---|---|---|---|---|---|
| 1 | University Ibadan | 359 | 7.4 | 423 | 61115 | 170.24 |
| 2 | Lagos State University | 355 | 7.3 | 312 | 6633 | 18.68 |
| 3 | Obafemi Awolowo University | 208 | 4.3 | 242 | 23575 | 113.34 |
| 4 | University Ilorin | 165 | 3.4 | 109 | 3382 | 20.50 |
| 5 | University Nigeria | 165 | 3.4 | 231 | 11253 | 68.20 |
| 6 | Lagos University | 164 | 3.4 | 74 | 2802 | 17.09 |
| 7 | Lagos University Teaching Hosp | 164 | 3.4 | 82 | 6781 | 41.35 |
| 8 | Nigerian Inst Med Res | 162 | 3.3 | 54 | 1613 | 9.96 |
| 9 | Ahmadu Bello University | 153 | 3.2 | 353 | 58165 | 380.16 |
| 10 | University Cape Town | 152 | 3.1 | 371 | 58361 | 383.95 |
| 11 | UCL | 137 | 2.8 | 423 | 62964 | 459.59 |
| 12 | University Coll Hosp | 136 | 2.8 | 249 | 45108 | 331.68 |
| 13 | University KwaZulu Natal | 136 | 2.8 | 352 | 52731 | 387.73 |
| 14 | University Washington | 120 | 2.5 | 382 | 66309 | 552.58 |
| 15 | University Benin | 117 | 2.4 | 131 | 14381 | 122.91 |
| 16 | University Oxford | 117 | 2.4 | 342 | 59210 | 506.07 |
| 17 | London Sch Hyg and Trop Med | 116 | 2.4 | 325 | 45202 | 389.67 |
| 18 | Addis Ababa University | 113 | 2.3 | 252 | 40511 | 358.50 |

| 19 | Ain Shams University | 113 | 2.3 | 370 | 58763 | 520.03 |
| 20 | Harvard University   | 110 | 2.3 | 356 | 57464 | 522.40 |

Their ACPP values (383.95 for the University of Cape Town, 459.59 for UCL, and 552.58 for the University of Washington) show that their joint research with the University of Lagos has made a significant global impact. The University of Oxford and Harvard University, two of the world's top academic institutions, have strong collaboration records with the University of Lagos, with TGCS of 59,210 and 57,464, respectively. Their ACPP values (506.07 for Oxford and 522.40 for Harvard) suggest highly influential research collaborations. Addis Ababa University and Ain Shams University represent African collaborations outside Nigeria, with high TGCS (40,511 for Addis Ababa and 58,763 for Ain Shams) and impressive ACPP values (358.50 and 520.03, respectively), indicating productive research partnerships. London School of Hygiene and Tropical Medicine (LSHTM), a globally renowned public health institution, has collaborated on 116 publications with the University of Lagos, resulting in an ACPP of 389.67, indicating strong contributions to health-related research.

This highlights both local (within Nigeria) and international collaborations, with the latter contributing significantly to the global impact of research from the University of Lagos. Institutions like the University of Washington, Harvard University, UCL, and Oxford have highly impactful collaborations, reflected in their high TGCS and ACPP. Local institutions like the University of Ibadan and Obafemi Awolowo University remain crucial partners, contributing heavily to the volume of research and making substantial local and global impacts. In summary, this reflects the broad range of collaborations the University of Lagos engages in, both within Nigeria and internationally. While local collaborations drive the majority of the research output, international partnerships contribute significantly to the global citation impact and the overall academic standing of the university's research.

**Analysis of Ranking of Department-wise Distribution**

This table 6 presents the research output of various departments and subdivisions of the University of Lagos. The departments are ranked based on the number of records, along with metrics like total local citation score (TLCS), total global citation score (TGCS), and average citations per paper (ACPP). These statistics reflect the research productivity and the impact of each department. College of Medicine leads in terms of research output, contributing to 29.9% of the university's publications (1,446 records). It has garnered a high TGCS of 23,735 and an ACPP of 16.41, showing that its research is influential both locally and globally. Teaching Hospital ranks second with 317 records, representing 6.6% of the total publications. It has a TGCS of 5,678 and an ACPP of 17.91, indicating its significant contribution to healthcare research. Department of Chemistry ranks third with 195 records (4%) and a strong ACPP of 27.53, reflecting that its research is highly cited and impactful with a TGCS of 5,368. The Faculty of Science has 192 records (4%) with an ACPP of 13.82. While its TGCS is lower compared to departments like Chemistry, its research is still well-recognized. The Department of Physics has contributed 162 publications, with a relatively high TLCS of 246 and TGCS of 2,022, resulting in an ACPP of 12.48. This shows its contributions to physics research are well-cited. The Faculty of Pharmacy produced 145 publications, but with a lower TGCS of 1,654 and an ACPP of 11.41, its overall citation impact is modest compared to other science-focused departments. The Department of Economics has 123 records, but its TGCS of 3,924 and ACPP of 31.90 make it a standout department in terms of the high impact of its research on economics and social sciences. The Department of Psychiatry has a relatively lower number of records (113), but it stands out for having the highest ACPP (528.19) and TGCS (59,685) in the entire list, indicating a very high level of global influence in its field. Department of Zoology and Department of Microbiology contribute a smaller percentage of research output, with 103 and 98 records, respectively, but their ACPP (9.76 for Zoology and 14.55 for Microbiology) shows moderate citation impact. The Department of Mechanical Engineering (93 records) and Faculty of Engineering (87 records) have TGCS of 1,482 and 894, respectively, showing their research is also contributing steadily to global scientific discourse, though their ACPP (15.94 and 10.28) indicates moderate impact. Department of Community Health and Primary Care is notable with an ACPP of 182.03, one of the highest in this table, indicating that its research, while fewer in number (69 records), has substantial influence globally (TGCS of 12,560). The Department of Chemical Engineering and Department of Mathematics show relatively lower citation impacts, with ACPPs of 10.83 and 7.30, respectively, despite contributing to the university's overall research productivity. The Department of Cell Biology and Genetics and the Department of Metallurgical and Materials Engineering contribute smaller records (68 and 67), but their citation impact is modest, as reflected by their TGCS and ACPP values.

**Table 6.** Ranking of department-wise distribution of research from University of Lagos.

| S.N. | Institution with sub-division | Records | % | TLCS | TGCS | ACPP |
|---|---|---|---|---|---|---|
| 1 | University of Lagos, College of Medicine | 1446 | 29.9 | 666 | 23735 | 16.41 |
| 2 | University of Lagos, Teaching Hospital | 317 | 6.6 | 122 | 5678 | 17.91 |
| 3 | University of Lagos, Department of Chemistry | 195 | 4 | 110 | 5368 | 27.53 |
| 4 | University of Lagos, Faculty of Science | 192 | 4 | 145 | 2653 | 13.82 |
| 5 | University of Lagos, Department of Physics | 162 | 3.3 | 246 | 2022 | 12.48 |
| 6 | University of Lagos, Faculty of Pharmacy | 145 | 3 | 42 | 1654 | 11.41 |
| 7 | University of Lagos, Department of Economics | 123 | 2.5 | 263 | 3924 | 31.90 |
| 8 | Lagos State University, College of Medicine | 121 | 2.5 | 74 | 1350 | 11.16 |
| 9 | University of Lagos, Department of Psychiatry | 113 | 2.3 | 297 | 59685 | 528.19 |
| 10 | Lagos University, Teaching Hospital | 109 | 2.3 | 51 | 2146 | 19.69 |
| 11 | University of Lagos, Department of Zoology | 103 | 2.1 | 33 | 1005 | 9.76 |
| 12 | University of Lagos, Department of Microbiology | 98 | 2 | 60 | 1426 | 14.55 |
| 13 | University of Lagos, Department of Mechanical Engineering | 93 | 1.9 | 38 | 1482 | 15.94 |
| 14 | University of Lagos, Faculty of Engineering | 87 | 1.8 | 25 | 894 | 10.28 |
| 15 | University of Lagos, Department of Botany and Microbiology | 76 | 1.6 | 91 | 1184 | 15.58 |
| 16 | University of Lagos, Department of Math | 74 | 1.5 | 21 | 540 | 7.30 |
| 17 | University of Lagos, Department of Chemical Engineering | 70 | 1.4 | 25 | 758 | 10.83 |
| 18 | University of Lagos, Department of Community Health and Primary Care | 69 | 1.4 | 113 | 12560 | 182.03 |
| 19 | University of Lagos, Department of Cell Biology and Genetics | 68 | 1.4 | 7 | 523 | 7.69 |
| 20 | University of Lagos, Department of Met and Mat Engn | 67 | 1.4 | 27 | 809 | 12.07 |

The College of Medicine and Teaching Hospital dominates in terms of research output and influence, particularly in healthcare and clinical research. The Department of Psychiatry stands out for having the highest average citations per paper, making it the most globally impactful department in terms of citation influence. STEM departments (Science, Technology, Engineering, and Mathematics) such as Chemistry, Physics, and Mechanical Engineering are highly productive and have strong ACPP values, highlighting their significance in global research. Social sciences (e.g., Economics) and health sciences (e.g., Community Health and primary Care) show high citation impact, indicating they contribute to important societal discussions and advancements in health. In summary, the research output from the University of Lagos is heavily driven by its medical, science, and engineering departments, with notable global citation impact from departments such as Psychiatry, Economics, and Chemistry.

**Distribution of Papers by Types of Documents**

This table 7 categorizes the different types of research documents produced by the University of Lagos and their corresponding contributions in terms of record count, percentage, total local citation score (TLCS), and total global citation score (TGCS). Articles dominate the research output, with 3,548 records (73.12% of total publications). These have the highest TLCS (2,636) and TGCS (123,005), reflecting that traditional articles are the most common and widely cited form of research from the university. Meeting Abstracts represent the second-largest category, with 787 records (16.22%), but have minimal citation impact (TLCS of 12 and TGCS of 139). This is expected, as meeting abstracts generally provide preliminary findings and tend to be less cited. Reviews comprise 275 records (5.67%), with TGCS of 10,111 and a notable TLCS of 118, indicating that review articles are highly impactful and widely referenced within the global research community. Editorial Material accounts for 67 records (1.38%), with modest citation impact (TLCS of 21 and TGCS of 1,101). Editorials generally have lower citation rates but still contribute to academic discourse. Letters (51 records, 1.05%) have slightly more citations than editorials, with TLCS of 23 and TGCS of 231. Letters typically reflect brief communications of research findings. Article; Early Access documents, with 38 records (0.78%), have no local citations yet but show a global citation count of 125. These papers are still in the early stage of publication and are likely to receive more citations as they gain visibility. Proceedings Papers account for 33 records (0.68%) and show TGCS of 862, indicating moderate citation activity for conference papers published as articles. Corrections (27 records, 0.56%) and Book Reviews (7 records, 0.14%) have negligible citation impact, as expected for these document types, which are not intended for original research contributions. Other document types, such as News Items, Biographical-Items, Art Exhibit Reviews, Retractions, and Retracted Publications, are represented by only one or a few records, contributing little to the university's overall citation metrics. These document types are rare and typically not central to research output.

**Table 7.** Document type contribution of research.

| S.N. | Document type | Records | % | TLCS | TGCS |
|---|---|---|---|---|---|
| 1 | Article | 3548 | 73.12 | 2636 | 123005 |
| 2 | Meeting Abstract | 787 | 16.22 | 12 | 139 |
| 3 | Review | 275 | 5.67 | 118 | 10111 |
| 4 | Editorial Material | 67 | 1.38 | 21 | 1101 |
| 5 | Letter | 51 | 1.05 | 23 | 231 |
| 6 | Article; Early Access | 38 | 0.78 | 0 | 125 |
| 7 | Article; Proceedings Paper | 33 | 0.68 | 12 | 862 |
| 8 | Correction | 27 | 0.56 | 0 | 1 |
| 9 | Book Review | 7 | 0.14 | 0 | 0 |
| 10 | Review; Early Access | 6 | 0.12 | 0 | 257 |
| 11 | News Item | 4 | 0.08 | 0 | 15 |
| 12 | Review; Book Chapter | 2 | 0.04 | 0 | 58 |
| 13 | Art Exhibit Review | 1 | 0.02 | 0 | 0 |
| 14 | Article; Data Paper | 1 | 0.02 | 0 | 3 |
| 15 | Article; Retracted Publication | 1 | 0.02 | 1 | 1 |
| 16 | Biographical-Item | 1 | 0.02 | 0 | 0 |
| 17 | Editorial Material; Early Access | 1 | 0.02 | 0 | 7 |
| 18 | Retraction | 1 | 0.02 | 0 | 0 |
| 19 | Review; Retracted Publication | 1 | 0.02 | 0 | 1 |
|  | *Total* | *4852* | *100* | *2823* | *135917* |

Articles are the core of the university's research contributions, both in terms of volume and citation impact. Meeting Abstracts and Reviews provide additional academic output, with reviews being especially impactful in terms of citations. Less common document types, such as Editorials, Letters, and Proceedings Papers, add diversity to the publication portfolio, but their citation impact is relatively modest. The presence of Corrections and Retractions indicates an active engagement with maintaining research integrity by correcting or retracting problematic publications when necessary. In summary, while Articles remain the primary vehicle for disseminating research from the University of Lagos, the inclusion of various document types contributes to the diversity and depth of its academic output.

**Most Relevant Countries by Corresponding Authors**

This table 8 presents the contributions of various countries to research articles associated with the University of Lagos, focusing on the number of articles, their percentage share, and metrics related to single-country publications (SCP) and multi-country publications (MCP). Nigeria is the predominant

**Table 8.** Most relevant countries by corresponding authors.

| S.N. | Country | Articles | Articles % | SCP | MCP | MCP % |
|---|---|---|---|---|---|---|
| 1 | Nigeria | 2683 | 55.3 | 1903 | 780 | 29.1 |
| 2 | USA | 374 | 7.7 | 1 | 373 | 99.7 |
| 3 | United Kingdom | 250 | 5.2 | 2 | 248 | 99.2 |
| 4 | South Africa | 130 | 2.7 | 0 | 130 | 100 |
| 5 | China | 91 | 1.9 | 0 | 91 | 100 |
| 6 | Germany | 73 | 1.5 | 0 | 73 | 100 |
| 7 | India | 40 | 0.8 | 1 | 39 | 97.5 |
| 8 | Canada | 39 | 0.8 | 0 | 39 | 100 |
| 9 | Netherlands | 37 | 0.8 | 0 | 37 | 100 |
| 10 | Australia | 35 | 0.7 | 0 | 35 | 100 |
| 11 | Benin | 24 | 0.5 | 22 | 2 | 8.3 |
| 12 | Italy | 20 | 0.4 | 0 | 20 | 100 |
| 13 | Malaysia | 17 | 0.4 | 1 | 16 | 94.1 |
| 14 | Ghana | 16 | 0.3 | 0 | 16 | 100 |
| 15 | Switzerland | 16 | 0.3 | 0 | 16 | 100 |
| 16 | Brazil | 14 | 0.3 | 0 | 14 | 100 |
| 17 | Finland | 12 | 0.2 | 0 | 12 | 100 |
| 18 | Saudi Arabia | 11 | 0.2 | 0 | 11 | 100 |
| 19 | Kenya | 9 | 0.2 | 0 | 9 | 100 |
| 20 | Korea | 9 | 0.2 | 0 | 9 | 100 |

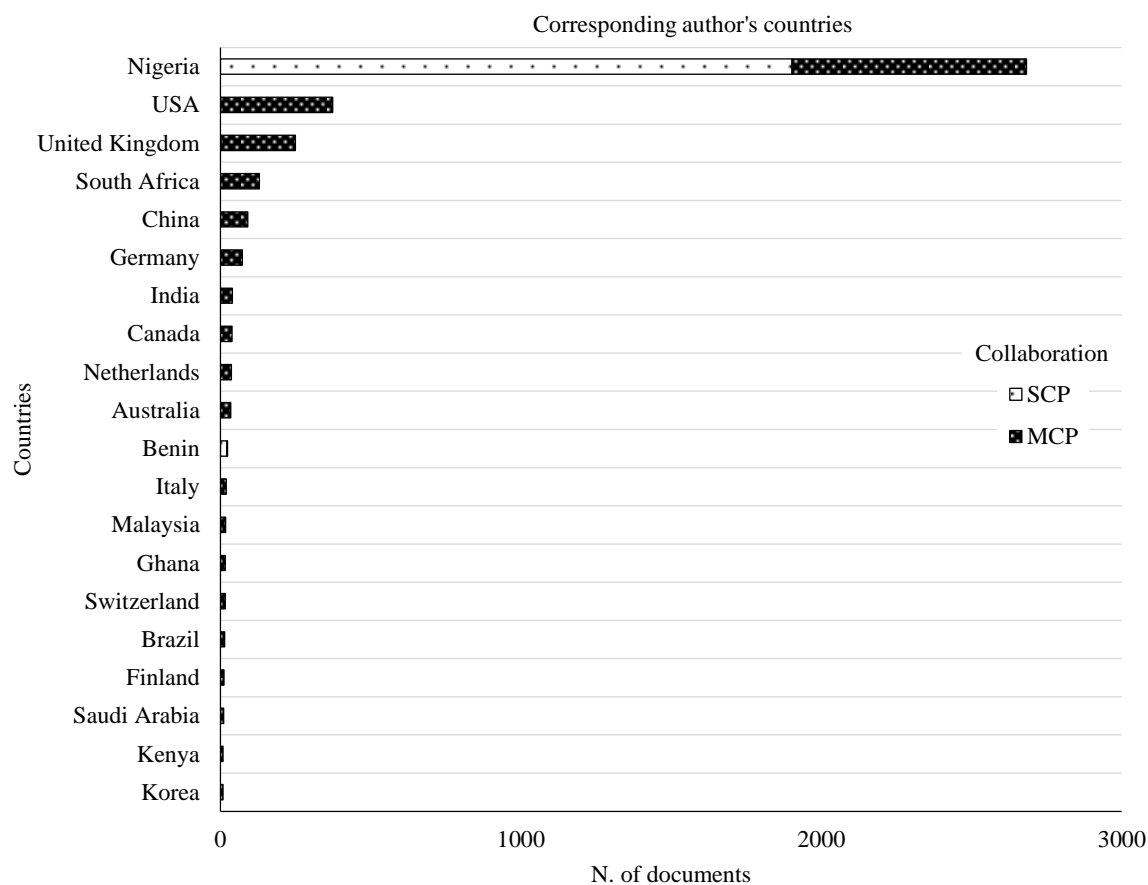

**Figure 3.** Corresponding author's countries . (SCP: Single country publications, MCP: multiple country publication) contributor, with 2,683 articles (55.3% of the total). (Source: web of science)

This indicates that a significant majority of research output comes from local authors, reflecting the university's strong national presence in research. Figure 3 The USA ranks second with 374 articles (7.7%). The vast majority (99.7%) of these articles are multi-country publications (373), indicating extensive collaboration with Nigerian authors. The United Kingdom follows closely with 250 articles (5.2%), also showing a high percentage of multi-country collaborations (99.2%). South Africa has 130 articles (2.7%), entirely in the form of multi-country publications, suggesting a strong collaborative relationship with Nigerian researchers. China contributes 91 articles (1.9%), all of which are also multi-country publications, indicating a focus on collaborative research. Germany (73 articles, 1.5%), India (40 articles, 0.8%), and Canada (39 articles, 0.8%) have similar patterns of strong multi-country collaboration with very few single-country publications. Benin, with 24 articles (0.5%), shows a unique distribution, with 22 being single-country publications (8.3% MCP). This indicates that most of the research from Benin is conducted independently, with limited international collaboration. Other countries like Italy, Malaysia, Ghana, and Switzerland show complete reliance on multi-country collaborations, further emphasizing the collaborative nature of modern research. Brazil, Finland, Saudi Arabia, Kenya, and Korea contribute minimally, each with fewer than 15 articles, and all relying solely on multi-country collaborations.

*High level of collaboration:* The data reveals that many of the countries with the highest article counts engage primarily in multi-country research collaborations, highlighting a trend towards international cooperation in academic research. *Dominance of local contributions:* Nigeria's substantial article count underscores the university's significant role in producing research outputs within the country. *Limited single-country publications:* Most countries listed, aside from Nigeria and Benin, show a tendency to collaborate internationally, which may enhance research quality and dissemination but also point to a possible dependency on collaborative networks for research output. In summary, the University of Lagos exhibits a robust national research output with extensive international collaborations, reflecting both local strength and global engagement in the research landscape.

**DISCUSSION**

The analysis presented highlights the significant contributions of researchers from the University of Lagos, Nigeria, to the global academic landscape. The data reveals trends in publication output, author impact, institutional collaboration, departmental contributions, document types, and the countries of corresponding authors. The findings indicate that the University of Lagos boasts a diverse array of prolific authors, with Mokdad AH leading in both h-index and citation metrics. This showcases the quantity of research output and its quality and impact on the academic community. The high citation scores and h-index values suggest that the research produced is widely recognized and has contributed meaningfully to the respective fields. The top sources preferred by researchers, predominantly journals like Lancet and Environmental Science and Pollution Research, underline the focus on high-impact and interdisciplinary research areas. Furthermore, the collaboration with other notable institutions, such as the University of Ibadan and Lagos State University, signifies a strong network of research partnerships that enhance the quality and reach of the studies conducted. This collaborative spirit is essential in addressing complex global challenges, such as health and environmental issues. The data indicates a rich distribution of research across various departments, particularly within the College of Medicine and the Teaching Hospital, suggesting a strong emphasis on health-related research. The significant contributions from fields such as Chemistry and Physics reflect the university's commitment to fostering a multidisciplinary approach to research, which is essential for innovative solutions. The overwhelming majority of publications are articles that demonstrate the university's commitment to producing substantial and rigorous research, with a notable emphasis on peer-reviewed articles that contribute to the body of knowledge. The presence of various document types, including meeting abstracts and reviews, indicates a dynamic research environment that encourages dissemination and discussion of findings. The prominence of Nigeria as the leading country in terms of article contributions emphasizes the importance of local research. However, the strong representation of international collaborators from countries like the USA and the UK reveals a trend of global academic integration. This collaboration not only enriches the research output but also enhances the visibility and impact of Nigerian researchers in the global arena. To enhance research output, the university should prioritize funding for high-impact research areas and establish clear incentives for interdisciplinary collaborations. By fostering an environment that encourages innovation and cross-disciplinary projects, the university can significantly improve its research visibility and impact.

**CONCLUSION**

This research is significant for several reasons. First, it provides a comprehensive overview of the research output from the University of Lagos, highlighting the institution's role as a leading center for academic excellence in Nigeria and beyond. By documenting the patterns of publication, collaboration, and author impact, this study serves as a vital resource for policymakers, educational institutions, and researchers aiming to understand the landscape of academic research in Nigeria. Furthermore, the findings underscore the importance of fostering international collaborations, as they enhance the quality and impact of research efforts. This is particularly relevant in today's interconnected world, where tackling global challenges such as pandemics and climate change necessitates collective efforts across borders. The significance of this research extends to other institutions and stakeholders in the academic community, providing insights into best practices for enhancing research visibility and impact. It can serve as a benchmark for other universities in Nigeria and Africa as they strive to increase their research output and reputation on the global stage.

Policymakers should invest in research infrastructure and encourage global partnerships. University administrators are advised to enhance the university's global presence and reward high-impact research, while researchers should focus on multidisciplinary projects and publishing in high-impact journals.

In conclusion, the research contributes to a greater understanding of the dynamics of academic research at the University of Lagos and underscores the necessity of collaboration, innovation, and strategic focus in advancing knowledge and addressing societal challenges. It lays the groundwork for future studies that can further explore these themes and their implications for the academic community at large.